\documentclass[sigconf, 
nonacm
]{acmart}

\usepackage{booktabs} %

\usepackage{setspace}

\usepackage{dirtree}
\usepackage{fontawesome}
\usepackage{subcaption}
\usepackage{longtable}
\usepackage{pdflscape}
\usepackage{graphicx}
\usepackage{multirow}
\usepackage{pgfplots}
\usepackage{pifont}
\usepackage{textcomp}
\usepackage{xcolor}
\usepackage{xspace}
\usepackage{url} 
\usepackage{hhline}
\usepackage{listings}

\newcommand{\frameworkname}[0]{Structured-GraphRAG\xspace}

\usepackage{colortbl}  %
\usepackage{xcolor}    %
\usepackage{hhline}    %

\usepackage[acronym]{glossaries}
\newacronym{ann}{ANN}{Approximate Nearest Neighbor}
\newacronym{asr}{ASR}{Automatic Speech Recognition}
\newacronym{ir}{IR}{Information Retrieval}
\newacronym{nlp}{NLP}{Natural Language Processing}
\newacronym{rag}{RAG}{Retrieval Augmented Generation}
\newacronym[plural=LLMs,firstplural=Large Language Models(LLMs)]{llm}{LLM}{Large Language Model}
\newacronym{kg}{KG}{Knowledge Graph}

\usepackage[utf8]{inputenc} %

\usepackage{hyperref}

\pgfplotsset{compat=1.18}

\begin{document}

\title{Enhancing Structured-Data Retrieval with GraphRAG: Soccer Data Case Study
}

\author{Zahra Sepasdar}
\affiliation{%
  \institution{Forzasys \& SimulaMet}
  \city{Oslo}
  \country{Norway}}
\email{zahra.sepasdar@gmail.com}

\author{Sushant Gautam}
\affiliation{%
  \institution{SimulaMet \& OsloMet }
  \city{Oslo}
  \country{Norway}}
\email{sushant@simula.no}

\author{Cise Midoglu}
\affiliation{%
  \institution{SimulaMet}
  \city{Oslo}
  \country{Norway}}
\email{cise@simula.no}

\author{Michael A. Riegler}
\affiliation{%
  \institution{SimulaMet \& OsloMet}
  \city{Oslo}
  \country{Norway}}
\email{michael@simula.no}

\author{Pål Halvorsen}
\affiliation{%
  \institution{SimulaMet, Forzasys \& OsloMet}
  \city{Oslo}
  \country{Norway}}
\email{paalh@simula.no}


\begin{abstract}
Extracting meaningful insights from large and complex datasets poses significant challenges, particularly in ensuring the accuracy and relevance of retrieved information. Traditional data retrieval methods such as sequential search and index-based retrieval often fail when handling intricate and interconnected data structures, resulting in incomplete or misleading outputs. To overcome these limitations, we introduce \frameworkname, a versatile framework designed to enhance information retrieval across structured datasets in natural language queries. \frameworkname utilizes multiple knowledge graphs, which represent data in a structured format and capture complex relationships between entities, enabling a more nuanced and comprehensive retrieval of information. This graph-based approach reduces the risk of errors in language model outputs by grounding responses in a structured format, thereby enhancing the reliability of results. We demonstrate the effectiveness of \frameworkname by comparing its performance with that of a recently published method using traditional retrieval-augmented generation. Our findings show that \frameworkname significantly improves query processing efficiency and reduces response times. While our case study focuses on soccer data, the framework’s design is broadly applicable, offering a powerful tool for data analysis and enhancing language model applications across various structured domains.
\end{abstract}


\maketitle
\section{Introduction and Related Work}\label{sec:introduction}
As digital content continues to expand rapidly, the need for advanced retrieval systems has become more critical. Users increasingly prefer to interact with these systems through natural language queries—posing questions as they would to another person—making it essential for retrieval technologies to understand and process language in a human-like manner~\cite{gautam2022assisting, Gautam2023Oct}. Large language models (LLMs) are integral to these systems for interpreting user queries accurately. However, LLMs face several challenges that impact their ability to retrieve relevant information effectively.
Firstly, these models often struggle to fully comprehend the context embedded within questions, which can lead to responses that do not align precisely with the user’s intentions. Furthermore, language models encounter challenges related to "hallucinating" information, i.e., generating entirely fictional facts. Another issue is that the training datasets may not be updated regularly, potentially leading to incorrect outputs. To address these challenges, one effective approach is using Retrieval-Augmented Generation (RAG), which integrates the advantages of retrieval-based methods to access up-to-date information with generation-based models for coherent and context-aware responses. This hybrid method helps mitigate issues related to outdated data and enhances the model's ability to provide accurate and relevant information.

To further advance beyond the traditional RAG framework, Graph\-RAG introduces a novel enhancement by integrating knowledge graphs (KGs). By leveraging the structured relationships and rich semantics within knowledge graphs, GraphRAG not only improves the retrieval process but also enables more nuanced and contextually aware responses. This incorporation of KGs allows the model to understand complex queries better and provide more detailed and interconnected information, thereby pushing the boundaries of what RAG frameworks can achieve.

 KG is a structured representation of knowledge that captures relationships between data points within a specific domain. This structure is composed of nodes that represent entities and edges that depict the relationships between these entities, forming an interconnected information. KGs are effective tools for organizing and integrating vast amounts of data from various sources, enhancing the accessibility and usability of this information. They play a crucial role in improving search engines by providing structured, interconnected search results, thereby offering users more relevant and comprehensive answers~\cite{Shaoxiong,Xiieee}. Moreover, KGs are essential in advancing artificial intelligence applications, including recommendation systems and natural language processing, by supplying a robust framework for understanding context and semantics. By serving as a structured source of information, KGs help reduce hallucinations in LLMs, leading to more accurate and reliable outputs~\cite{Ba,Ci}.

Unlike RAG, which primarily focuses on retrieving relevant data without explicitly modeling the connections between pieces of information, GraphRAG uses KGs to capture the relationships and dependencies among data points. This approach allows for a deeper understanding of the retrieved information, leading to more accurate and context-aware responses. By structuring knowledge in a graph format, GraphRAG can reason over interconnected data, providing richer insights and more coherent outputs~\cite{Yu}. This method involves a retrieval module that sources additional data from external repositories, creating a "grounding context" that is integrated into the LLM's prompt. By incorporating this context, the model can generate responses that are not only more accurate but also more relevant to specific queries~\cite{Yun, SepasdarIronGraph}. In GraphRAG, the integration of KGs with LLMs significantly boosts their accuracy and reliability, ensuring that the outputs are both precise and trustworthy~\cite{Xia}.

It is worth noting that while RAG systems based on SQL are highly capable, the advantages of using a graph-based structure are: more accurate responses and reduced retrieval times. In fields such as digital content creation, news generation, and chatbots (like the current study) the ability to provide correct answers is crucial, particularly because language models can sometimes generate hallucinations.
One of the key strengths of a graph-based approach is that it adds a greater level of detail to the dataset. This additional layer enables better data representation and helps mitigate hallucinations by improving the model's ability to find relevant and accurate connections. Furthermore, graph structures make it easier to identify nodes and their related edges, which streamlines data retrieval and results in shorter response times. The combination of improved accuracy and faster response times is a major reason for the growing interest in graph-based systems for language models. These systems are particularly effective in applications where accuracy and timeliness are critical, providing a robust solution to some of the limitations inherent in more traditional retrieval methods.

Applying GraphRAG models requires the expertise of a graph theory specialist to construct KGs tailored to each dataset. In this paper, we introduce a novel framework called \frameworkname, specifically designed to enhance data retrieval across various structured datasets. Structured data refers to datasets that adhere to a predefined format, which can be efficiently managed and queried using SQL. Our approach presents a novel method for automatically generating KGs from structured datasets. This innovation removes the need for specialized, enabling a broader range of users to analyze data or build chatbots from structured data without requiring deep expertise in graph theory. Theroperty is the key advancement of our \frameworkname framework.
To demonstrate the capabilities of \frameworkname, we applied it to the soccer data from the SoccerNet dataset~\cite{SoccerNet}. By constructing KGs from this dataset, \frameworkname improves both the precision and ease of process in natural language queries. The results of our experiments show that compared to traditional RAG methods applied to the same dataset~\cite{soccerrag}, \frameworkname offers several key advantages. Notably, the structured nature of KGs significantly reduces the incidence of hallucinations in LLMs, thereby enhancing the reliability and accuracy of the outputs. Moreover, the graph-based architecture of our framework facilitates faster data retrieval, optimizing query performance and reducing response times. While our initial demonstration of \frameworkname’s effectiveness is based on soccer data, the framework is designed to be adaptable to any structured dataset, making it a versatile tool for a broad range of applications. 

A key challenge with existing graph-based approaches, such as GraphRAG, is that they often require a domain expert to meticulously design the KGs tailored to the specific data source to achieve optimal results. Without this specialized input, these methods may not perform better than non-graph-based approaches, limiting their utility. To address this limitation in data analysis, our paper provides a comprehensive explanation of how to systematically construct graphs from structured data sources. We propose a novel method for generating KGs from structured datasets, using the SoccerNet dataset as a practical example. This method not only simplifies the process of creating graphs but also demonstrates the broader applicability of our approach to a wide range of structured data. By providing a clear methodology for graph construction, we aim to make advanced data retrieval and analysis accessible to a wider audience, beyond those with specific graph expertise.

The structure of the paper is as follows:
we introduce \frameworkname in Section~\ref{sec:frameworkoverview}.
Section~\ref{sec:dataset} describes our source dataset. In Section~\ref{sec:framework}, we explain how to create KGs for structured data. In Section~\ref{sec:example}, our framework is illustrated with a comprehensive example. Section~\ref{sec:evaluation} presents evaluations of the framework performance. Section \ref{sec:discussion} discusses the weaknesses and strengths of the method, and Section~\ref{sec:conclusion} concludes the paper.

\section{Introduction to \frameworkname}\label{sec:frameworkoverview}
This section describes the methodology of our framework and its operation in response to user queries. To begin using the framework, the initial step is to create KGs from the source dataset. A complete explanation of the KGs structures is provided in Section~\ref{sec:framework}. For now, let's assume the KGs have been created, and we want to utilize them. These KGs are designed to provide accurate and detailed information from the source dataset and are stored in a graph database such as Neo4j. In this research, we employed the Neo4j Python driver to establish and manage connections with the Neo4j database, enabling efficient query execution and manipulation of graph data. This integration is part of our Python-based analytical framework.

\begin{figure}[h]
    \centering
    \vspace{-3mm}
    \includegraphics[width=0.9\columnwidth]{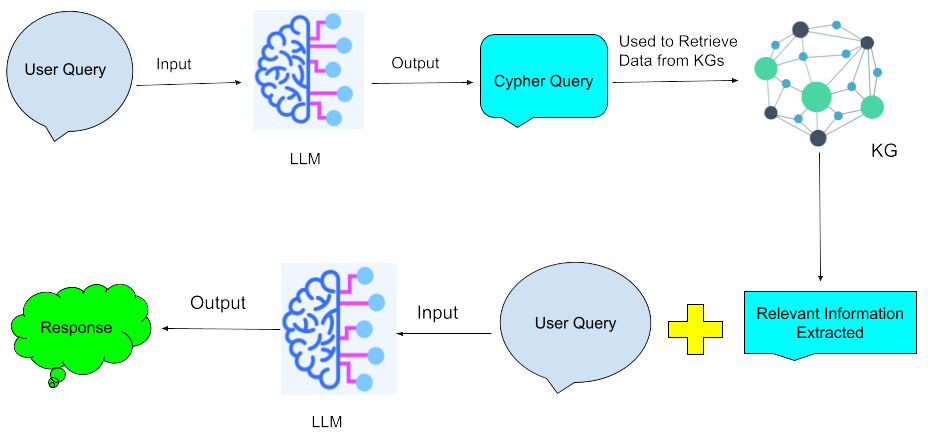}
    \caption{Framework overview.}  %
    \label{fig:pgraphrag}  %
    \vspace{-4mm}
\end{figure}

Once the KGs are set up, the system is prepared for user interaction. When a user asks a question, \frameworkname processes this query using an LLM. The LLM translates the user's natural language query into a Cypher query, which is a specialized query language for interacting with graph databases. The Cypher query is then used by a smart search tool, which navigates through the graph database to identify relevant nodes and edges, effectively pinpointing the most pertinent pieces of information from the constructed KGs. This process allows the system to extract specific data points and relationships that directly address the user’s query. After retrieving this data, the system combines it with the context of the user’s original question and re-inputs it into the LLM. The LLM utilizes this combined information to generate a comprehensive and nuanced response, providing the user with a detailed answer that is both accurate and informative. Figure~\ref{fig:pgraphrag} illustrates the complete process, from query to the final response. 

In summary, our framework operates through a series of steps designed to maximize the efficiency and accuracy of information retrieval and response generation:

\hspace{0.01cm} \textbf{Step 1. KGs Construction:} The framework begins by constructing KGs from the dataset, which are then stored within a graph database.

\hspace{0.01cm} \textbf{Step 2. Query Translation:} When a user submits a query, the system translates it into Cypher queries to interact with the graph database effectively.

\hspace{0.01cm} \textbf{Step 3. Information Retrieval:} These Cypher queries are employed to navigate the graph database, extracting relevant nodes and edges that provide the necessary information.

\hspace{0.01cm} \textbf{Step 4. Answer Generation:} Finally, the LLM uses the retrieved graph data, combined with the context of the original user query, to generate a detailed and accurate response.

This methodology leverages advanced LLMs and graph technology to develop a dynamic system that provides precise, relevant, and comprehensive answers to user queries about dataset sources. It is important to clarify that we use OpenAI's GPT-3~\cite{OpenAIGPT3} and GPT-4~\cite{OpenAIGPT4} because other emerging models, such as Llama 2~\cite{touvron2023llama,Llama-2} and Mistral-7B~\cite{jiang2023mistral}, lack the advanced function-calling capabilities necessary for integrating language models with external tools. While fine-tuning these models might improve their functionality, it is a resource-intensive and time-consuming process. Therefore, GPT-3 and GPT-4 provide a more reliable and efficient solution for the objectives of this research.

\section{Showcase Study: Soccer Data}\label{sec:dataset}
To demonstrate the performance and effectiveness of our framework, we use soccer data from SoccerNet~\cite{SoccerNet,SoccerNetv2}, an extensive dataset designed for the analysis and understanding of soccer videos. Table~\ref{tab:dataset-summary} summarizes the source dataset, detailing sub-datasets, divisions, corresponding files, and data types. In this section, we provide a comprehensive description of the data structure to enhance understanding of the KGs construction process. This detailed explanation ensures the methodology behind the graph's architecture is transparent and reproducible, facilitating further research.

\begin{table}[h]

    \scriptsize
    \centering
    
    \begin{tabular}{|c|c|l|l|}
    
        \hline
        \multicolumn{2}{|c|}{\textbf{Source Dataset}} 
        & \multirow{2}{*}{\textbf{Corresponding File(s)}} 
        & \multirow{2}{*}{\textbf{Type}} 
        \\ \hhline{--~~} 

        \textbf{Sub-dataset}
        & \textbf{Division} 
        & 	
        & 
        \\ \hline 

Labels 
        & - 
        & \texttt{Labels-v2.json}  
        & Structured 
        \\ \hline 

 \multirow{2}{*}{Captions} 
        & Players
        & \texttt{Labels-caption.json} 
        & Structured 
        \\ \hhline{~---} 
        & Annotations
        & \texttt{Labels-caption.json}   
        & Unstructured 
        \\ \hline 

    \end{tabular}
    \caption{Summary of the source dataset.}
    \label{tab:dataset-summary}
    
\end{table}

\begin{figure}[h]
    \centering
    \newcommand{\figw}{0.6\columnwidth}
    \fbox{%
        \begin{minipage}{\figw}
            \centering
            \includegraphics[width=\columnwidth]{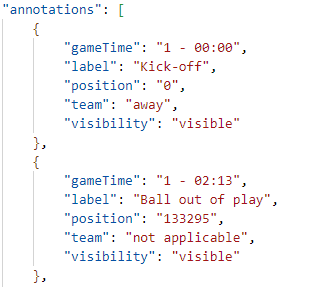}
        \end{minipage}%
    }
    \caption{Samples of data in Labels file.}
    \label{fig:labeldataset}
\end{figure}

\textbf{Label Dataset:}
The structure of the data in the Labels file is shown in Figure~\ref{fig:labeldataset}.

\begin{figure}[h]
    \centering
    \newcommand{\figw}{0.65\columnwidth}
    \fbox{%
        \begin{minipage}{\figw}
            \centering
            \includegraphics[width=\columnwidth]{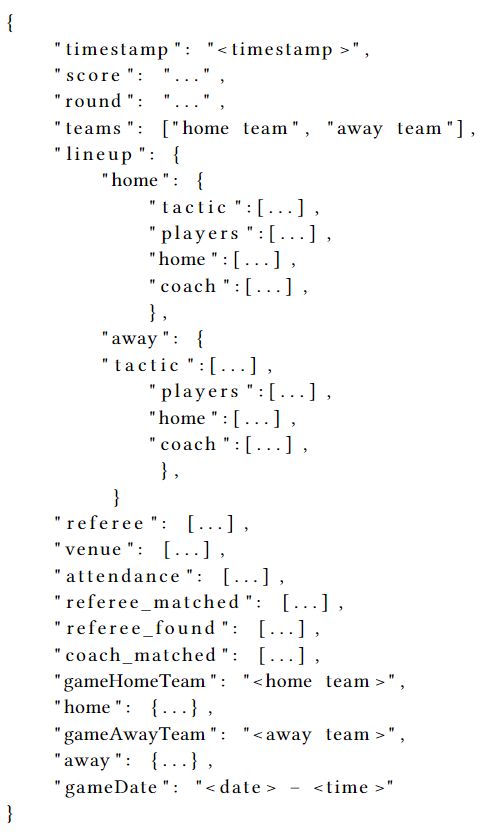}
        \end{minipage}%
    }
    \caption{Captions file.}
    \label{fig:captiondataset}
\end{figure}

\begin{figure}[h]
    \centering
    \newcommand{\figw}{0.65\columnwidth}
    \fbox{%
        \begin{minipage}{\figw}
            \centering
            \includegraphics[width=\columnwidth]{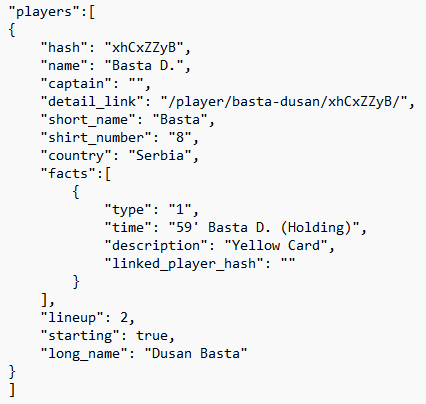}
        \end{minipage}%
    }
    \caption{Samples of Players Data.}
    \label{fig:p}
\end{figure}

\textbf{Caption Dataset:}
The structure of data in the Captions file is shown in Figure~\ref{fig:captiondataset}.
The format of the player data in Caption file (Line 8 of Figure~\ref{fig:captiondataset}) is depicted in Figure~\ref{fig:p}.
As you see in Figure~\ref{fig:p}, fact of player data consist of different types. These are categorized into seven types, as detailed in
Table~\ref{tab:event_types} (no fact of type 5).

\begin{table}[h]
    \centering
    \resizebox{0.31\textwidth}{!}{ %
    \begin{tabular}{|c|l|}
    \hline
    \textbf{Type} & \textbf{Description} \\
    \hline
    1 & Relates to yellow cards \\
    \hline
    2 & Relates to red cards \\
    \hline
    3 & Relates to goals \\
    \hline
    4 & Relates to own goals \\
    \hline
    6 \& 7 & Related to substitutions \\
    \hline
    8 & Indicates who assisted the goal \\
    \hline
    \end{tabular}
    }
    \caption{Description of Fact Types.}
    \label{tab:event_types}
\end{table}

\section{KG construction}\label{sec:framework}
In this section, we explore a novel approach to transforming structured datasets into KGs. Leveraging the SoccerNet dataset as a primary example, we demonstrate how this methodology can be generalized to any dataset that is organized in a tabular format. By converting structured data into a KG, we aim to provide a flexible framework that reveals the complex relationships and deeper insights within the data, which are often not immediately apparent in traditional formats. This approach not only facilitates enhanced data visualization and analysis but also opens up new possibilities for uncovering meaningful connections across diverse domains.

For each game, we work with two separate datasets: Labels and Captions-Players. We generate a unique KG for each of these datasets to capture the distinct information they contain.

\subsection{Game \& Team Nodes}
\begin{figure}[htbp]
    \centering
    \newcommand{\figw}{0.69\columnwidth}
    \begin{subfigure}[b]{\figw}
        \centering
        \includegraphics[width=\columnwidth]{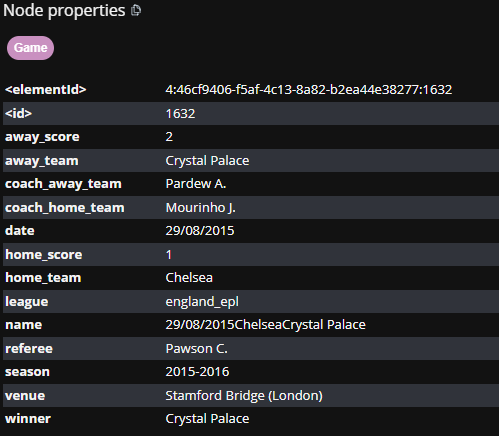}
    \end{subfigure}
  \caption{An example of Game node.}
  \label{fig:game-node}
\end{figure}

\begin{figure}[htbp]
    \centering
    \newcommand{\figw}{0.69\columnwidth}
    \begin{subfigure}[b]{\figw}
        \centering
        \includegraphics[width=\columnwidth]{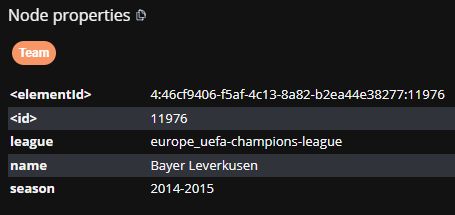}
    \end{subfigure}
  \caption{An example of Team node.}
  \label{fig:team-node}
\end{figure}

Each game in the dataset is represented as a node, labeled as \texttt{Game}, with its attributes obtained from Figure 3. An example of such a node and its attributes is illustrated in Figure~\ref{fig:game-node}, which shows the Chelsea vs. Crystal Palace game.
Each team is represented by a node with two attributes league, and season. Figure~\ref{fig:team-node} shows an example of a team node for Bayer Leverkusen.

\begin{figure}[h]
    \centering
    \newcommand{\figw}{0.9\columnwidth}
    \begin{subfigure}[b]{\figw}
        \centering
        \includegraphics[width=\columnwidth]{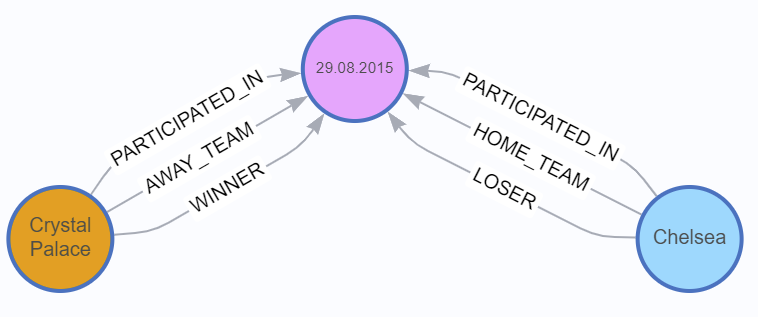}
    \end{subfigure}
  \caption{An example of Game and Team connections.}
  \label{fig:gameteam}
\end{figure}

Each team is connected to the corresponding game with an edge named \texttt{PARTICIPATED\_IN}. Additionally, each team is connected to the game by an edge named \texttt{HOME\_TEAM} or \texttt{AWAY\_TEAM}, depending on whether the team is hosting. If the result of the game is not a draw, then two additional edges named \texttt{WINNER} and \texttt{LOSER} are added. Refer to Figure~\ref{fig:gameteam} to see the connections between the game and the teams for the Chelsea vs. Crystal Palace game.

\subsection{Labels KG Construction}\label{sec:knowledge-graph}
\begin{figure}[htbp]
    \centering
    \newcommand{\figw}{0.69\columnwidth}
    \begin{subfigure}[b]{\figw}
        \centering
        \includegraphics[width=\columnwidth]{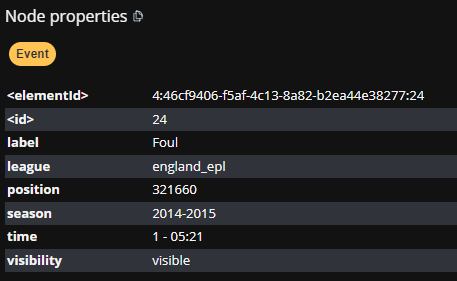}
    \end{subfigure}
  \caption{An example of Label node.}
  \label{fig:label-node}
\end{figure}

Each entry in the Label file represents a node with the label \texttt{Event}, where the node's name is specified by data.label. For example, Figure~\ref{fig:label-node} illustrates a node created for a foul event along with its attributes.

\begin{figure}[h]
    \centering
    \newcommand{\figw}{0.5\columnwidth}
    \begin{subfigure}[b]{\figw}
        \centering
        \includegraphics[width=\columnwidth]{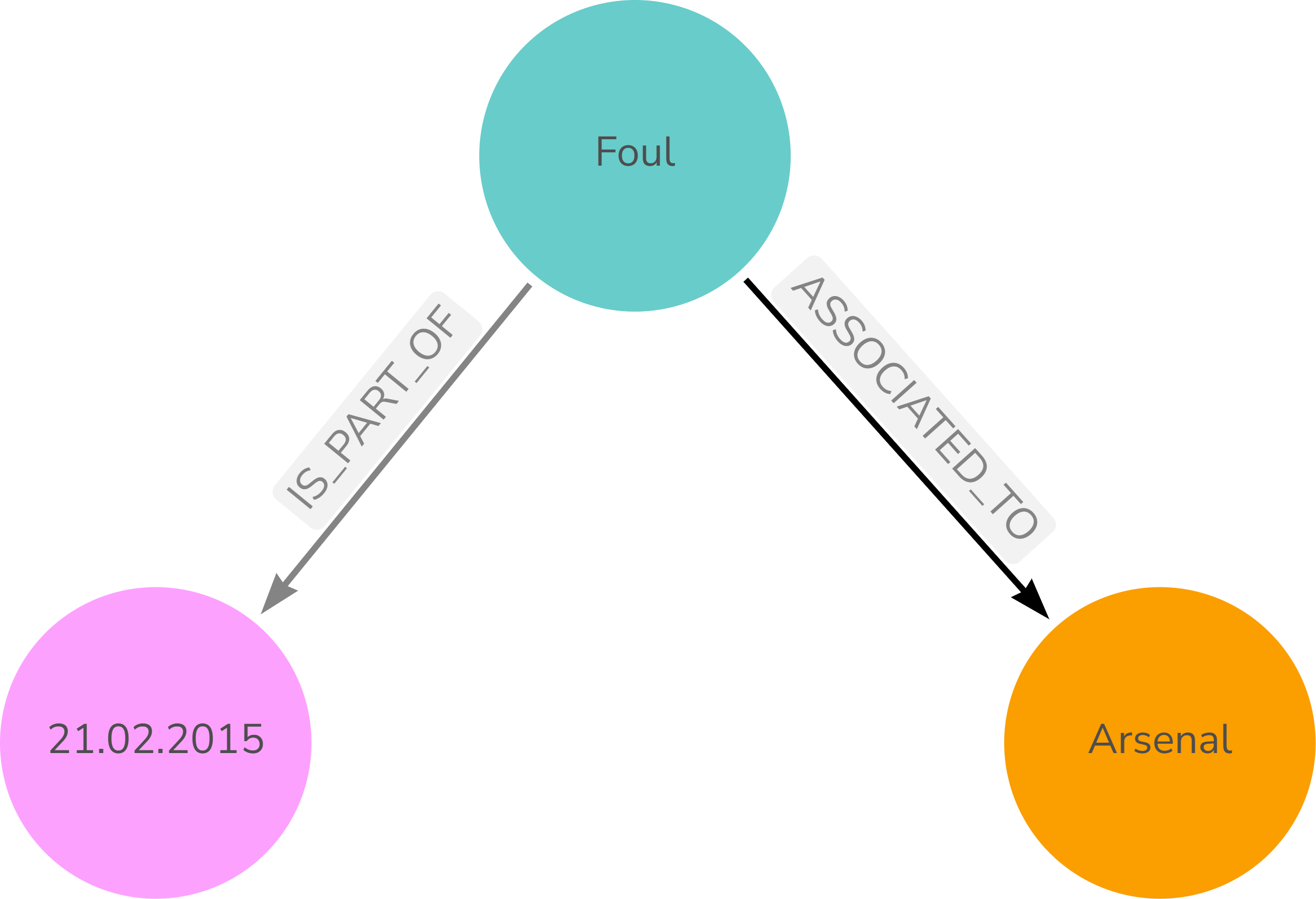}
    \end{subfigure}
  \caption{An example of Event connections.}
  \label{fig:eventteamconnection}
\end{figure}

As shown in Figure~\ref{fig:labeldataset}, data.team  can take various categorical values.
If data.team is not ’not applicable’, the event node is connected
to the corresponding team by an edge named\texttt{ ASSOCIATED\_TO}. Each event node is connected to the correspondent game node with the edge named \texttt{IS\_PART\_OF}. Figure~\ref{fig:eventteamconnection} illustrates the connections of the event node for Arsenal mach. 

\subsection{Captions KG Construction}
\begin{figure}[h]
    \centering
    \newcommand{\figw}{0.69\columnwidth}
    \begin{subfigure}[b]{\figw}
        \centering
        \includegraphics[width=\columnwidth]{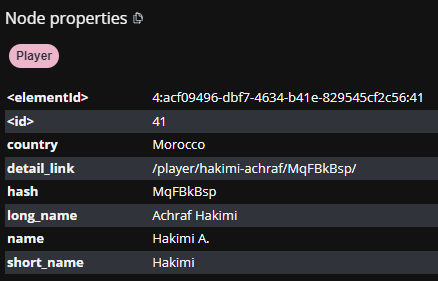}
    \end{subfigure}
  \caption{An example of a Player node.}
  \label{fig:playernode}
\end{figure}

\begin{figure}[h]
    \centering
    \newcommand{\figw}{0.5\columnwidth}
    \begin{subfigure}[b]{\figw}
        \centering
        \includegraphics[width=\columnwidth]{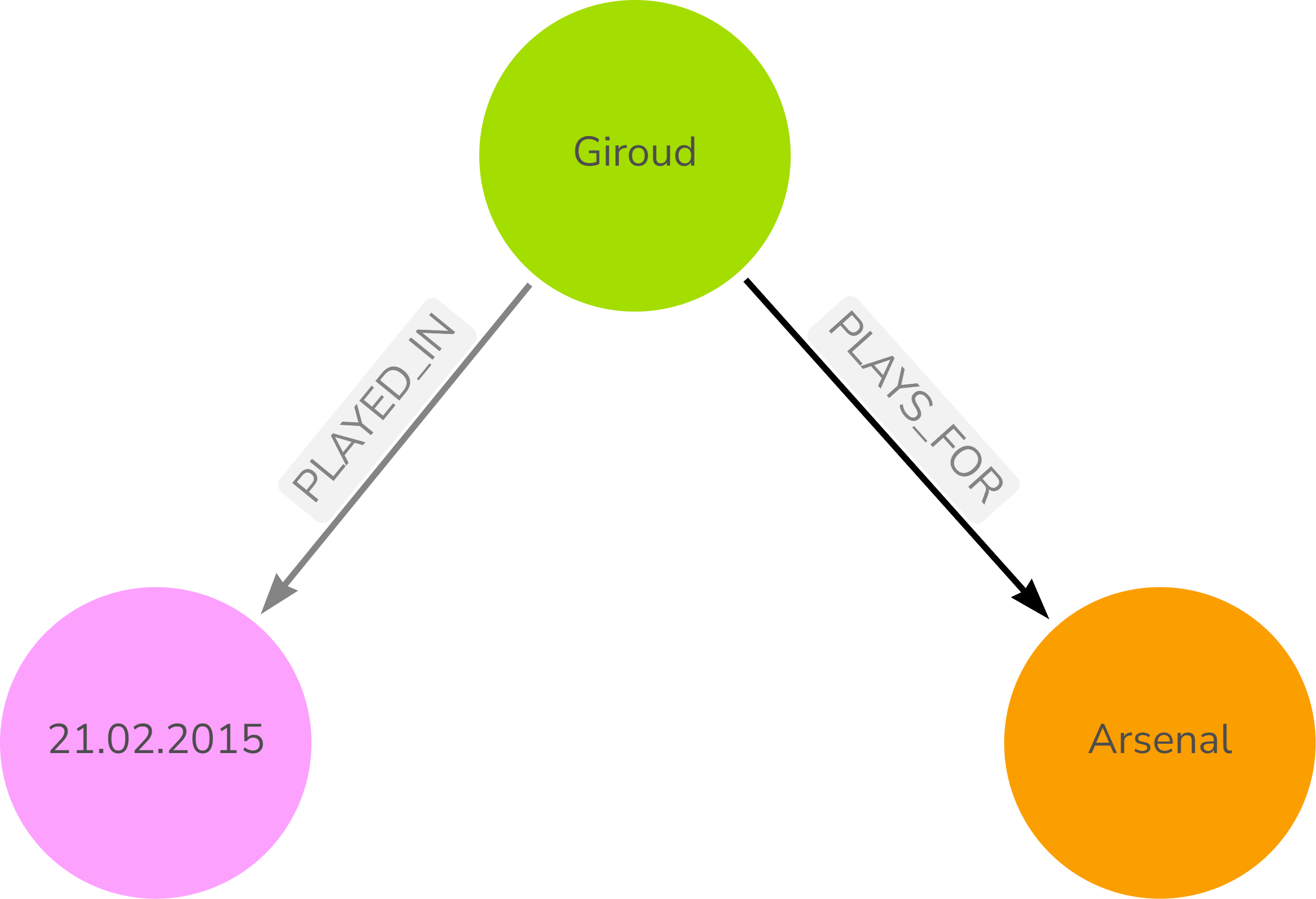}
    \end{subfigure}
  \caption{An example of a Player node connections.}
  \label{fig:playerconnection}
\end{figure}

For each player, a node with label \texttt{Player} is created.
An example of a player node and its attributes is illustrated in Figure~\ref{fig:playernode}.
Each player is connected to the corresponding game by an edge named \texttt{PLAYED\_IN}. Each player is connected to the corresponding team by an edge named \texttt{PLAYS\_FOR}. An example of such connections are  shown in Figure~\ref{fig:playerconnection}.
If the player has data in the fact part, we create additional nodes. As shown in Table~\ref{tab:event_types}, there are different facts types. To explain the construction of the graph, we describe the method for one of theseFfct s and do not detail the others, as they can be created similarly.

\begin{figure}[h]
    \centering
    \newcommand{\figw}{0.69\columnwidth}
    \begin{subfigure}[b]{\figw}
        \centering
        \includegraphics[width=\columnwidth]{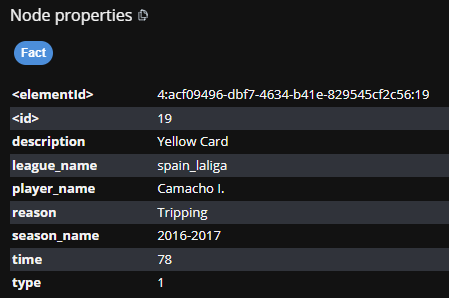}
    \end{subfigure}
  \caption{An example of Fact node with Type 1.}
  \label{fig:yellowcardnode}
\end{figure}

\begin{figure}[h]
    \centering
    \newcommand{\figw}{0.59\columnwidth}
    \begin{subfigure}[b]{\figw}
        \centering
        \includegraphics[width=\columnwidth]{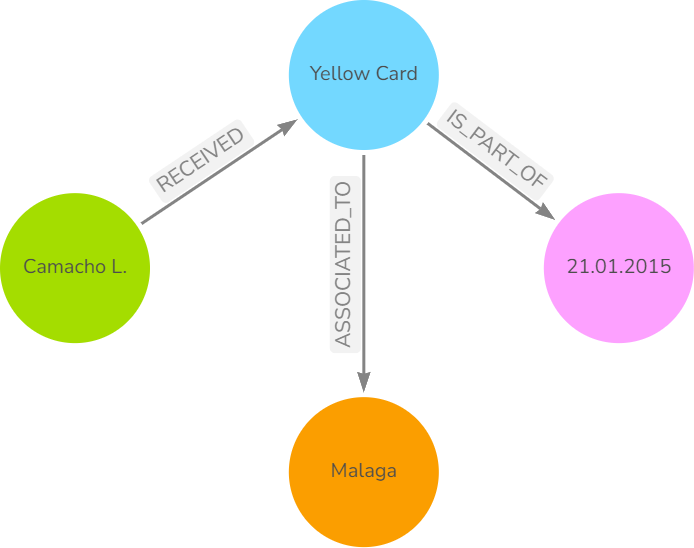}
    \end{subfigure}
  \caption{All connections for Yellow Card node.}
  \label{fig:fact1}
\end{figure}

The structure of fact with Type 1 is presented in Figure~\ref{fig:p}. As illustrated, "time" consists of three elements: first, the exact time events occurred; second, the name of the player who received the yellow card, and third the reason why the player received the card.
We create one node for Yellow Card with label \texttt{Fact}. An example of a Fact node with Type 1 and its attributes is illustrated in Figure~\ref{fig:yellowcardnode}.
Each \texttt{Fact} node is connected to the corresponding game by an edge named \texttt{IS\_PAR\_OF}, to the corresponding team by an edge named \texttt{ASSOCIATED\_TO}, and to the player node by an edge named \texttt{RECEIVED}. These connections are shown in Figure~\ref{fig:fact1}.

Even though we detailed the process of constructing KGs using the SoccerNet dataset as a case study, the methodology is not limited to this specific dataset; it is a versatile approach that can be applied to any structured dataset, typically organized in a tabular format. The core idea is to transform the structured data into a KG by selecting one column (or row) to represent the nodes and using the remaining columns (or rows) to define the attributes of these nodes. The edges, or the connections between nodes, are then defined based on the specific relationships pertinent to the dataset's domain. For instance, in a healthcare dataset, edges might represent relationships such as "diagnosed with," "prescribed," or "related to a symptom," reflecting the unique connections within that context. This method of converting structured data into a KG allows for a tailored representation of data, where the semantics of the edges are thoughtfully chosen to convey the most meaningful relationships.

This approach enables the transformation of structured data into a KG that captures the underlying complexities and interrelations inherent in the data. This not only enhances data visualization and analysis but also unveils hidden patterns and insights that may remain obscured in traditional tabular formats. Ultimately, this method provides a powerful tool to derive information from structured datasets across various domains.

\section{Detailed Example}\label{sec:example}
To demonstrate retrieval of information from KG structures, we provide a comprehensive example to illustrate how our framework functions in practice and how it leverages the interconnected nature of KGs to efficiently extract relevant data. By doing so, we aim to highlight the advantages of using graph-based approaches for knowledge representation and retrieval, including enhanced query capabilities, improved data integration, and more intuitive exploration of complex datasets.

\textbf{Example:} Suppose a user submits the following query: "Give me the total home goals for Bayern Munich in the 2014-15 season."

\begin{figure}[h]
    \centering
    \frame{\includegraphics[trim={0 0 0cm 1cm},clip,width=1.0\columnwidth,]{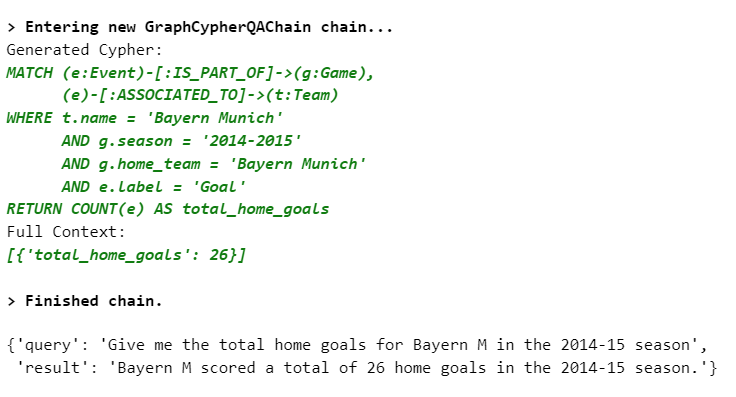}}
    \caption{Sample of a  Q\&A application.}
    \label{fig:q1}
\end{figure}

\begin{itemize}
     \item \textbf{Step 1.} The KGs as described in Section \ref{sec:framework} are created from the SoccerNet dataset and stored in Neo4j. It is important to mention that the KGs creation from the dataset occurs only once, and when using the framework, we actually begin from Step 2.
    \item \textbf{Step 2.} To extract relevant information from our KGs, the first task is to translate the user query into a format that the KGs can process. This involves using an LLM to convert the natural language query into a Cypher query, a specialized language for querying graph databases. In this Cypher query, elements of the user query are mapped to nodes and edges within the KGs. For instance, 'goal' is treated as an 'Event' node, 'Bayern Munich' as a 'Team' node, and the relationship between them as an 'ASSOCIATED\_TO' edge. The Cypher query then searches for a 'Game' node with the attributes 'season' set to 2014-2015 and 'home\_team' as Bayern Munich, which is linked to an 'Event' node via an 'IS\_PART\_OF' edge. This process is illustrated in Figure \ref{fig:q1} (Generated Cypher).
    \item \textbf{Step 3.} The Cypher query generated in Step 1 is executed against the KG, retrieving the relevant nodes and edges. The details of this execution and the retrieved data are provided in the ‘Full Context’ section of Figure \ref{fig:q1}.
    \item \textbf{Step 4.} Finally, the initial user query, along with the retrieved data from the 'Full Context,' are fed into the LLM. Using this combined information, the LLM generates the answer, which is shown in the ‘Result’ section of Figure \ref{fig:q1}.
\end{itemize}

\section{Evaluation}\label{sec:evaluation}
In this section, we evaluate the performance of \frameworkname and provide an in-depth analysis. Table \ref{tab:example_questions} highlights a sample set of questions derived from the dataset. It is important to note that the examples listed are not exhaustive, there could be many more examples for each category.  For the purposes of demonstrating the effectiveness of our approach, we have selected a subset of 10 questions.

\begin{table}[h!]
\centering
    \small
    \resizebox{\columnwidth}{!}{
    \begin{tabular}{|p{1cm}|p{5cm}|p{6cm}|}
        \hline
        \textbf{Q \#} & \textbf{Category} & \textbf{Example} \\ \hline
        Q1 & Questions about existence of a team/player in a dataset: & Is Manchester United in the database? \\ \hline
        Q2 & Questions about the total home events for a team in a specific season: & Give me the total home goals for Bayern Munich in the 2014-15 season. \\ \hline
        Q3 & Questions about the home team advantage in a season: & Calculate the home advantage for Real Madrid in the 2015-16 season. \\ \hline
        Q4 & Questions about the number of goals that a player scored in a specific season/league: & How many goals did Messi score in the 2015-16 season? \\ \hline
        Q5 & Questions about the number of yellow cards/red cards that a player received in a specific season/league: & How many yellow cards did Enzo Perez get in the 15-2016 season? \\ \hline
        Q6 & Questions about all teams that played against a specified team in a particular season and league: & List all the teams that played a game against Napoli in the 2016-17 season in Serie A. \\ \hline
        Q7 & Questions about all teams in a specific league, in a specific season: & Give all the teams in the league UCL in the 2015-2016 season. \\ \hline
        Q8 & Questions to obtain information from events in the first/second half of a season/league: & Give me all games in the EPL with yellow cards in the first half in the 2015-2016 season. \\ \hline
        Q9 & Questions to obtain information related to teams of a player: & What teams and leagues has Adnan Januzaj played in? \\ \hline
        Q10 & Questions about the number of goals in a specific game in a season/league: & How many goals did E. Hazard score in the game between Bournemouth and Chelsea in the 2015-2016 season? \\ \hline
        Q11 & Questions to obtain detailed information about goals in a specific game within a season or league: & Who assisted Marcelo with goals in the 2015-16 season? \\ \hline
        Q12 & Questions about the reason for yellow cards/red cards that a player received in a specific season/league: & Why did Enzo Perez get yellow cards in the 2015-2016 season? \\ \hline
        Q13 & Questions about the number of yellow cards/red cards in a specific season/league: & How many yellow and red cards were given in the UEFA Champions League in 2015-2016, and did anyone receive a red card? \\ \hline
        Q14 & Questions to obtain detailed information about events in a specific game within a season or league: & Which events happened in the game between Bournemouth and Chelsea in the 2015-2016 season? \\ \hline
        Q15 & Questions about the dates when a specific event happened in a given league during a particular season: & Make a list of when corners happened in the English Premier League (EPL) 2015-2016 season. \\ \hline
        Q16 & Questions to obtain information about a player: & What information do you have about Cristiano Ronaldo in your dataset? \\ \hline
        Q17 & Questions about referees: & Tell me about the games that Atkinson M. was the referee for. \\ \hline
        Q18 & Questions about coaches: & List the names of the coaches for Arsenal. \\ \hline
        Q19 & Questions about venues: & Which games are played at Santiago Bernabéu Stadium? \\ \hline
        Q20 & Questions about game information: & Tell me all information about the game between Chelsea and Burnley in 2014-2015. \\ \hline
        Q21 & Questions about determining which team won more games in a season: & Which team won more games in the 2014-15 season, Arsenal or Liverpool? \\ \hline
        Q22 & Questions about the number of games a team won in a season: & How many games did Arsenal win in the 2014-15 season? \\ \hline
        Q23 & Questions about leagues that involved several teams: & What league is Manchester United, Arsenal, Bournemouth, Real Madrid, Chelsea, and Liverpool in? \\ \hline
    \end{tabular}}
    \caption{Examples of Different Types of Questions.}
    \label{tab:example_questions}
\end{table}

\subsection{Optimal KGs}
One of the main advantages of applying graph-based methods is that the data processing time is significantly shorter compared to other methods. However, this only occurs when the KGs structures created from the source dataset are optimal. In this context, an optimal graph is one that contains all relevant information from the dataset while being organized in a way that minimizes complexity. This efficient structure ensures that the system can quickly identify important nodes and edges without requiring extensive computation. 

In Section \ref{sec:framework}, we explained how the KGs are designed to comprehensively represent all the information in the dataset. However, it is also crucial to demonstrate that these structures, while being highly informative, are not excessively complex. To illustrate the simplicity of these graphs, we use the concept 
density from graph theory. 

\textbf{Definition:} Suppose that $G=(V,E)$ is a directed graph, where $\left| V \right|$
 represents the number of nodes, and $\left| E \right|$
 represents the number of edges. The density of $G$ is shown by $D(G)$ and defined as the ratio of the number of directed edges to the maximum possible number of directed edges.
\begin{align*}
D(G) = \frac{\left|E\right|}{\left|V\right| \times (\left|V\right| - 1)}.
\end{align*}
The density of a directed graph ranges from 0 to 1, where a lower density indicates a graph with fewer edges relative to the number of possible edges, making it simpler and less computationally demanding to manage. This simplicity is particularly advantageous when working with large datasets, as it allows for faster processing and retrieval of information. In our framework, the low densities of the Labels KG ($2.85 \times 10^{-6}$) and the Captions KG ($4.47 \times 10^{-4}$) indicate that these graphs are sparse, meaning they have very few edges compared to the maximum number of possible edges. Such sparsity is crucial for efficiency because it reduces the computational overhead required to traverse the graph and locate relevant nodes and edges. The sparsity of the Label and Caption KGs ensures that our framework can quickly identify and retrieve the necessary information without being bogged down by an excessive number of irrelevant connections. Therefore, our designed KGs structure significantly enhances the system’s performance, enabling it to handle large-scale data efficiently.

\subsection{Execution Time}\label{sec:t}
As previously discussed, our KGs are optimal, leading to a substantial reduction in execution time. To illustrate this reduction, we conduct a performance comparison with a recently published paper~\cite{soccerrag}, which utilized the same dataset as our study. The key distinction between the two approaches lies in how the data is processed: while the authors of ~\cite{soccerrag} analyze the dataset in its raw form, our method first transforms the data into KGs before performing analysis. This comparison clearly demonstrates the superior performance of our KG-based approach in reducing execution time compared to the direct data analysis method employed in ~\cite{soccerrag}.

For the evaluation, we applied both methods to a series of sample queries and measured their execution times. The results are presented in Table \ref{tab:query-response-time}, where the column indicating percentage improvement clearly shows that our KG-based approach consistently delivers faster response times compared to the method used in ~\cite{soccerrag}. These results highlight the efficiency gains achieved by our framework, demonstrating its ability to process and respond to queries with significantly greater speed, thereby confirming the superior performance of \frameworkname over the direct data analysis approach employed in ~\cite{soccerrag}.

\begin{table}[h!]
\centering
\renewcommand{\arraystretch}{1.5} %
\resizebox{\linewidth}{!}{%
\begin{tabular}{|c|c|c|c|}
\hline
\textbf{Query} & \textbf{Time (Method in \cite{soccerrag})} & \textbf{Time (\frameworkname)} & \textbf{Percentage Improvement (\%)} \\ \hline
Q1 & 4.61 & 1.66 & 64.00\% \\ \hline
Q2 & 22.23 & 4.00 & 82.01\% \\ \hline
Q3 & 595.98 & 6.16 & 98.97\% \\ \hline
Q4 & 94.74 & 2.48 & 97.38\% \\ \hline
Q5 & 49.47 & 3.40 & 93.13\% \\ \hline
Q6 & 109.46 & 3.90 & 96.44\% \\ \hline
Q7 & 56.97 & 3.80 & 93.33\% \\ \hline
Q8 & 106.33 & 4.80 & 95.48\% \\ \hline
Q9 & 63.60 & 3.90 & 93.87\% \\ \hline
Q10 & 59.52 & 6.90 & 88.41\% \\ \hline
\end{tabular}%
}
\caption{Execution Time Comparison Between Method in \cite{soccerrag} and \frameworkname.}
\label{tab:query-response-time}
\end{table}

\subsection{Accuracy}
For the accuracy evaluation, we designed an experiment to measure the consistency of the language model's responses. We posed the set of ten questions to the model five times consecutively, focusing on whether it consistently generated the correct answers across all iterations. This repetitive questioning approach is crucial for evaluating how well the model handles hallucination, which is a common issue in LLMs. The result of our experiments are indicated in Table \ref{tab:accuracy}.

The accuracy percentage is obtained using the formula in Equation \ref{eq
}: \begin{equation} \label{eq
} \text{Accuracy} = \left( \frac{\text{number of correct answers}}{\text{total number of questions}} \right) \times 100. \end{equation} As we have 10 questions, each asked 5 times, this results in a total of 50 questions. According to Table \ref{tab:accuracy}, there are 18 correct answers when applying the method from \cite{soccerrag}, resulting in 36\% accuracy, and 32 correct answers when applying \frameworkname, resulting in 64\% accuracy. Our experimental results demonstrate that the graph-based approach we employed significantly improves accuracy compared to the method used in ~\cite{soccerrag}. This highlights the effectiveness of integrating graph-based techniques to reduce hallucinations and provide more consistent, accurate outputs across multiple iterations.

\begin{table}[ht!]
    \centering
    \begin{tabular}{|c|c|c|c|c|c!{\vrule width 2pt}c|c|c|c|c|}
        \hline
        \multirow{3}{1.3cm}{\textbf{Query $\downarrow$ Iteration$\rightarrow$}} 
        & \multicolumn{5}{|c!{\vrule width 2pt}}{\textbf{Method in~\cite{soccerrag}}}
        & \multicolumn{5}{|c|}{\textbf{\frameworkname}}
        \\ \hhline{~----------} 
        & 1 & 2 & 3 & 4 & 5 & 1 & 2 & 3 & 4 & 5 \\ [0.2cm]\hline
        
        Q1 & \textcolor{green}{\ding{51}} 
           & \textcolor{green}{\ding{51}} 
           & \textcolor{green}{\ding{51}} 
           & \textcolor{green}{\ding{51}} 
           & \textcolor{green}{\ding{51}} 
           & \textcolor{green}{\ding{51}} 
           & \textcolor{green}{\ding{51}} 
           & \textcolor{green}{\ding{51}} 
           & \textcolor{green}{\ding{51}} 
           & \textcolor{green}{\ding{51}} \\ \hline

        Q2 & \textcolor{red}{\ding{55}} 
           & \textcolor{red}{\ding{55}} 
           & \textcolor{green}{\ding{51}} 
           & \textcolor{red}{\ding{55}} 
           & \textcolor{red}{\ding{55}} 
           & \textcolor{green}{\ding{51}} 
           & \textcolor{green}{\ding{51}} 
           & \textcolor{red}{\ding{55}} 
           & \textcolor{green}{\ding{51}} 
           & \textcolor{green}{\ding{51}} \\ \hline

        Q3 & \textcolor{red}{\ding{55}} 
           & \textcolor{green}{\ding{51}} 
           & \textcolor{red}{\ding{55}} 
           & \textcolor{red}{\ding{55}} 
           & \textcolor{green}{\ding{51}} 
           & \textcolor{red}{\ding{55}} 
           & \textcolor{green}{\ding{51}} 
           & \textcolor{red}{\ding{55}} 
           & \textcolor{green}{\ding{51}} 
           & \textcolor{green}{\ding{51}} \\ \hline

        Q4 & \textcolor{red}{\ding{55}} 
           & \textcolor{red}{\ding{55}} 
           & \textcolor{red}{\ding{55}} 
           & \textcolor{red}{\ding{55}} 
           & \textcolor{green}{\ding{51}} 
           & \textcolor{green}{\ding{51}} 
           & \textcolor{green}{\ding{51}} 
           & \textcolor{green}{\ding{51}} 
           & \textcolor{green}{\ding{51}} 
           & \textcolor{green}{\ding{51}} \\ \hline

        Q5 & \textcolor{red}{\ding{55}} 
           & \textcolor{red}{\ding{55}} 
           & \textcolor{red}{\ding{55}} 
           & \textcolor{red}{\ding{55}} 
           & \textcolor{red}{\ding{55}} 
           & \textcolor{green}{\ding{51}} 
           & \textcolor{green}{\ding{51}} 
           & \textcolor{red}{\ding{55}} 
           & \textcolor{green}{\ding{51}} 
           & \textcolor{green}{\ding{51}} \\ \hline

        Q6 & \textcolor{red}{\ding{55}} 
           & \textcolor{red}{\ding{55}} 
           & \textcolor{red}{\ding{55}} 
           & \textcolor{red}{\ding{55}} 
           & \textcolor{green}{\ding{51}} 
           & \textcolor{red}{\ding{55}} 
           & \textcolor{red}{\ding{55}} 
           & \textcolor{red}{\ding{55}} 
           & \textcolor{green}{\ding{51}} 
           & \textcolor{red}{\ding{55}} \\ \hline

           Q7 & \textcolor{green}{\ding{51}} 
           & \textcolor{red}{\ding{55}} 
           & \textcolor{green}{\ding{51}} 
           & \textcolor{red}{\ding{55}} 
           & \textcolor{green}{\ding{51}} 
           & \textcolor{green}{\ding{51}} 
           & \textcolor{red}{\ding{55}} 
           & \textcolor{green}{\ding{51}} 
           & \textcolor{green}{\ding{51}} 
           & \textcolor{red}{\ding{55}} \\ \hline

           Q8 & \textcolor{red}{\ding{55}} 
           & \textcolor{red}{\ding{55}} 
           & \textcolor{red}{\ding{55}} 
           & \textcolor{red}{\ding{55}} 
           & \textcolor{red}{\ding{55}} 
           & \textcolor{red}{\ding{55}} 
           & \textcolor{green}{\ding{51}} 
           & \textcolor{red}{\ding{55}} 
           & \textcolor{red}{\ding{55}} 
           & \textcolor{green}{\ding{51}} \\ \hline

           Q9 & \textcolor{red}{\ding{55}} 
           & \textcolor{green}{\ding{51}} 
           & \textcolor{red}{\ding{55}} 
           & \textcolor{green}{\ding{51}} 
           & \textcolor{red}{\ding{55}} 
           & \textcolor{green}{\ding{51}} 
           & \textcolor{red}{\ding{55}} 
           & \textcolor{green}{\ding{51}} 
           & \textcolor{red}{\ding{55}} 
           & \textcolor{red}{\ding{55}} \\ \hline

           Q10 & \textcolor{red}{\ding{55}} 
           & \textcolor{green}{\ding{51}} 
           & \textcolor{red}{\ding{55}} 
           & \textcolor{green}{\ding{51}} 
           & \textcolor{green}{\ding{51}} 
           & \textcolor{red}{\ding{55}} 
           & \textcolor{green}{\ding{51}} 
           & \textcolor{green}{\ding{51}} 
           & \textcolor{red}{\ding{55}} 
           & \textcolor{green}{\ding{51}} \\ \hline
    \end{tabular}
    \caption{Comparison of accuracy between Method in~\cite{soccerrag} and \frameworkname.}
    \label{tab:accuracy}
\end{table}

\section{Discussion}\label{sec:discussion}
In designing \frameworkname, a key feature is its ability to identify and correct inaccuracies in team names, league names, or player names caused by user queries. This functionality is crucial as it significantly reduces the occurrence of errors associated with retrieving incorrect data due to input mistakes.

Within the Player-Facts KG structure, some players are represented both as attributes within event nodes and as separate nodes connected to these events. This design aims to minimize hallucinations that can occur when generating Cypher queries with LLMs.

The KGs developed here are dynamic, allowing for updates with new data and adaptation as new information becomes available regarding game events, players, teams, coaches, referees, and stadiums. Additionally, this method for constructing KG structures is flexible and adaptable to various structured datasets, making it suitable for creating KG structures from different sources.

The \frameworkname framework presented in this paper is adaptable to various datasets and can be utilized for querying data from other datasets. To demonstrate its effectiveness, we applied it to the SoccerNet dataset~\cite{SoccerNet}. 

As illustrated in Table \ref{tab:accuracy}, \frameworkname outperforms the method presented in [17] on Questions 4 and 5. This highlights the advantage of employing a graph-based approach, where the smart search efficiently locates nodes with the desired attributes. Consequently, this reduces processing time and enhances answer accuracy. However, for tasks involving listing, the performance of our framework is comparable to that of the method in [17].
One of the challenges in \frameworkname using LLM lies in
handling questions that ask for lists. In such cases, while the LLM
can generate the correct Cypher query and retrieve all relevant
answers, it often fails to present the complete list.

Our results in Figure \ref{fig:q1} differ from those obtained through a Google search. This discrepancy arises because our dataset is SoccerNet, and we have only analyzed games from this specific dataset.

\section{Conclusion}\label{sec:conclusion}
In this paper, we introduced \frameworkname, a novel framework designed to enhance data retrieval across various structured datasets by utilizing the capabilities of KGs and graph-based architectures. Our results demonstrate that \frameworkname not only improves precision and reduces hallucinations in language models but also optimizes query performance and reduces response times. These advanced data retrieval techniques pave the way for applications in diverse fields, showcasing the versatility and adaptability of \frameworkname.

Additionally, by addressing the challenges of traditional graph-based methods, particularly the need for domain-specific expertise in KG design, we present a more accessible approach to graph construction for structured datasets, broadening its usability to a wider range of users.

\bibliographystyle{ACM-Reference-Format}
\bibliography{references}

\end{document}